\documentclass[aps,nofootinbib,twocolumn,prl,pacs,class-pre]{revtex4} 
\usepackage{setspace} \usepackage{amsmath,amssymb,amsfonts,amsthm}
\usepackage{graphicx}

\newcommand{\be}{\begin{equation}}
\newcommand{\ee}{\end{equation}}
\newcommand{\beq}{\begin{eqnarray}}
\newcommand{\eeq}{\end{eqnarray}}

\def\lsim{\hbox{ \raise.35ex\rlap{$<$}\lower.6ex\hbox{$\sim$}\ }}
\def\gsim{\hbox{ \raise.35ex\rlap{$>$}\lower.6ex\hbox{$\sim$}\ }}

\begin{document}
\title{Space-time dimensionality from brane collisions}
 \author{William Nelson\footnote{william.nelson@kcl.ac.uk},
         Mairi Sakellariadou\footnote{mairi.sakellariadou@kcl.ac.uk}}
\affiliation{Department of
  Physics, King's College, University of London, Strand WC2R 2LS,
  London, U.K.}

\begin{abstract}
Collisions and subsequent decays of higher dimensional branes leave
behind three-dimensional branes and anti-branes, one of which could
play the r\^ole of our universe.  This process also leads to the
production of one-dimensional branes and anti-branes, however their
number is expected to be suppressed.  Brane collisions may also lead
to the formation of bound states of branes. Their existence does not
alter this result, it just allows for the existence of one-dimensional
branes captured within the three-dimensional ones.

\end{abstract}
 

\maketitle

\section{Introduction}
The advent of string theory, and more recently non-commutative
geometry --- as an approach to unifying gravity with the other
fundamental forces --- has led to considerations of spatial dimensions
greater than three. If these theories are to have physical merit,
then they should provide a mechanism by which our, undoubtfully
$(3+1)$-dimensional, universe is produced. By achieving that we will
get, with the same token, an answer to the long-standing
puzzle of explaining the space-time dimensionality by purely
scientific means as opposed to (a strong form of the) anthropic
principle.

In string theory, following the Kaluza-Klein approach, space
dimensionality is typically explained by assuming that the extra
dimensions are tightly curled up, so as to be beyond our experimental
reaches. Precisely how this compactification occurs and gets
stabilised is a major difficultly in string theory that has yet to be
resolved. Within the Kaluza-Klein approach, the {\sl string gas
  scenario}~\cite{Brandenberger:1988aj} --- based on the target-space
duality relating string theories compactified on large and small
  tori, by interchanging winding and Kaluza-Klein states --- has been
  proposed long ago. According to this scenario, the universe started
  with all its dimensions being initially of the string scale,
  $\sqrt{\alpha'}$, while for dynamical reasons only three spatial
  dimensions were expanded. This can be easily understood, since only
  in a four-dimensional hyper-surface the world-sheets of winding and
  anti-winding modes can naturally overlap, and thus annihilate
  leading to the subsequent expansion of only three spatial
  dimensions. The string gas scenario has been supported by cosmic
  string experiments on a lattice~\cite{Sakellariadou:1995vk}.

It has also been noted that whilst critical superstring theory
requires $(9+1)$ dimensions --- at the basic level to ensure that the
fermionic world-volume degrees of freedom can be simultaneously Weyl
and Majorana --- in non-critical strings the dimensionality of
space-time is a dynamical parameter. Some progress has been made
recently towards understanding how four-dimensional non-critical
string theory behaves~\cite{Alexandre:2006xh}.

In light of large extra dimensions, it has been argued in
Ref.~\cite{Durrer:2005nz}, that starting with a distribution of branes
--- of any possible dimension allowed from the theory --- embedded in
a higher dimensional bulk, brane interactions could naturally lead to
the survival of only three-dimensional branes --- one of which could
play the r\^ole of our universe --- and one-dimensional branes
(D-strings) --- which could play the r\^ole of cosmic strings ---
called cosmic superstrings~\cite{Sakellariadou:2008ie}.  Strictly
speaking, in Ref.~\cite{Durrer:2005nz} only interactions between
branes of the same dimensionality were considered.  In what follows,
we repeat this argument, whilst extending it to allow for collisions
between branes of different dimension.  The collision of branes of
different dimension leads to complications because of the possibility
of forming bound states in which a D$p$-brane absorbs a D$q$-brane
(for $q<p$) to form a bound D$\left( p,q \right)$-brane system. Whilst
it has long been known that such a system is described by a D$p$-brane
with world-volume gauge fields ({\sl see}, for example
Ref.~\cite{johnson}), it is only recently that the tachyonic decay of
such branes --- due to either their non-BPS nature or the presence of
an anti-${\bar{\rm D}}p$-brane --- has been explicitly
derived~\cite{Nelson:2008eu}.

In this note, we derive the conditions under which two branes collide
and hence decay, or form a bound state composed of two branes of
unequal dimension. Our analysis is performed within the context of
type IIB string theory\footnote{By applying T-duality, one could
  equally well consider type IIA string theory.}. We show that brane
collisions are unlikely for D$p$-branes, with $p\leq3$, even if bound
states were formed during previous brane collisions.  We hence show
that given an initially random distribution of D$p$-branes --- where
the brane dimensionality $p$ is any odd number from 1 to 9 ---
embedded in a higher dimensional bulk ($d-1=9$, with $d-1$ the number
of spatial dimensions), the end point of the decay chain will
generically be D$3$-branes and possibly D$1$-branes (and their
anti-brane counterparts). We thus firstly confirm the argument of
Ref.~\cite{Durrer:2005nz}, and secondly address the issue raised in
Ref.~\cite{Karch:2005yz} regarding the r\^ole of bound states.  More
explicitly, we show that the possible existence of bound states does
not alter the conclusion of Ref.~\cite{Durrer:2005nz}, and thus, by a
simple {\sl geometric} argument, we address successfully the
origin of space-time dimensionality within the realm of string theory.

\section{Bound state decays}
Consider a D$p$-brane, with $p$ the number of spatial dimensions and
choose coordinates such that the brane would be aligned along
directions $\sigma_\mu = \left(1,2,\dots, p\right)$, if it were
flat. Excitations of the brane geometry are given by world-volume
scalars $X^m\left(\sigma_\mu\right)$, describing deviations from
flatness along the remaining $m$ (with $m=p+1,\dots,d-1$) directions.

The bosonic part of the Dirac-Born-Infeld (DBI) effective Lagrangian
for a D$p$-brane in terms of the U(1) gauge field strength and the
scalar fields $X^m$ reads
\be
 {\cal L}_{\rm eff} = - \sqrt{-\det\left( \eta_{\mu\nu}+\partial_\mu X^m
\partial_\nu X_m + F_{\mu\nu} \right)}~;
\ee
where $F_{\mu\nu}=\partial_\mu A_\nu - \partial_\nu A_\mu$ is the
world-volume electromagnetic field, with $A_\mu$ world-volume gauge
fields; $\eta_{\mu\nu}$ is the $(9+1)$-dimensional Minkowski
metric. Note that here we are neglecting the fermionic sector of the
D-brane action, however all the analysis below can be extended to
explicitly include fermions.

If the brane is non-BPS, then there is a tachyon present in its
spectrum. The dynamics of the tachyon field on a non-BPS D$p$-brane of
type IIA or IIB sueperstring theory is given from the tachyon effective
action~\cite{sen1999}
\beq
 {\cal L}_{\rm eff} &=& -V(T) \left[ -\det\left(
   \eta_{\mu\nu}+\partial_\mu X^m \partial_\nu X_m \right.\right.\nonumber
   \\  && \left.\left. ~~~~~~~~~~~~~~~~~~~~~
+ F_{\mu\nu} +\partial_\mu T \partial_\nu T\right)
   \right]^{1/2}~, 
\eeq
in terms of the massless gauge fields $A_\mu$ (with $0\leq \mu,\nu\leq
p$) and the transverse scalar fields $X^m$ (with $(p+1)\leq m\leq 9$)
on the world-volume of the non-BPS brane. The tachyon $T$ is a scalar
and $V\left(T\right)$ is the tachyon potential taken to be
non-negative. Note that $V(T)$ has a unique local maximum at the
origin ($T=0$) and a unique global minimum --- far from the origin ---
where the potential vanishes.

It has been shown~\cite{Sen:2003tm} that the tachyonic potential for a
non-BPS D$p$-brane contains an infinitely thin kink solution of finite
tension describing a co-dimension one BPS D-brane, as a topological
remnant of the bound states decay.  The action and dynamics of this
kink can be calculated and precisely agree with those of a
D$(p-1)$-brane. This remains true in the presence of world-volume
gauge fields~\cite{Nelson:2008eu}, provided the tachyonic kink is
aligned along the direction of the gauge fields.

If a D$q$-brane collides with a D$p$-brane, with $q<p$, then it
dissolves into the higher dimensional D$p$-brane, with its degrees of
freedom becoming gauge (magnetic) fields ({\sl see}, for example
Ref.~\cite{johnson}).  Since we are specifically looking at the case
where $q<p$, these magnetic fields will be aligned along
$q$-directions on the D$p$-brane, leaving $(p-q)$-directions (with
$(p-q)\geq 1$) available for the non-BPS D$\left(p,q\right)$-brane to
decay along. The result of this decay process will be either a
D$\left(p-1,q\right)$-brane, if $q<p-1$, or the same D$q$-brane that
was initially dissolved, if $q=p-1$.

It is important to note that the presence of such a dissolved brane
does not affect the decay mechanism~\cite{Nelson:2008eu}.  More
precisely, as we have shown in Ref.~\cite{Nelson:2008eu}, a bound
system composed by two branes of different dimensionality, D$p$, D$q$
with $q<p$, which can be described by just a D$p$-brane with
world-volume gauge fields, decays exactly as a standard D$p$-brane,
i.e., by forming a bound state of a D$(p-1)$ and a D$q$-brane. Thus,
the resulting defect is localised, as expected.

\section{Bound state collisions}
When a D$p$-brane collides with an anti-brane of the same dimension,
$\bar{\rm D}p$-brane, there is again a tachyon in the string
spectrum. If we take both branes to have the same transverse
directions, so that they are parallel, the low energy effective action
for this system reads~~\cite{sen1999}
\beq 
{\cal L}_{\rm eff} &=& -V\left( |T|, |X^m_{(1)} - X^m_{(2)}|\right)\nonumber\\
&&~~~~~~~~~\times \left[ \sqrt{
    -\det {\cal M}_{(1)} } + \sqrt{ -\det {\cal M}_{(2)} } \right]~,
\eeq 
where 
\beq
{\cal M}_{(\Delta)\mu\nu} &=& \eta_{\mu\nu} + \partial_\mu
X^m_{(\Delta)} \partial_\nu X_{(\Delta) m}+ F_{(\Delta) \mu\nu} \nonumber\\
&&+\frac{1}{2} (D_\mu T)^\star( D_\nu T) 
+\frac{1}{2}(D_\nu T)^\star( D_\mu T)~\nonumber
\eeq 
and
\beq
 F_{(\Delta) \mu\nu} =\partial_\mu A_{(\Delta)\nu}-\partial_\nu
 A_{(\Delta)\mu}\nonumber
\eeq
\beq
D_\mu T = \left( \partial_\mu - i A_{(1)\mu} + i
A_{(2)\mu}\right)T
\eeq
with $\Delta = 1,2$. The potential, $V$, depends only on the magnitude
of the tachyon $|T|$ --- which in this case is complex --- and $\sum_m
\left[ X^m_{(1)} + X^m_{(2)} \right]^2$.  Clearly for $T=0$, the above
action reduces to the sum of the DBI action on the individual branes.

As in the non-BPS case, the potential has been shown~\cite{Sen:2003tm}
to contain topological obstructions to reaching the true vacuum, with
the defect this time being of co-dimension $2$. This vortex-like
defect again has precisely the same dynamics as a D$(p-2)$-brane. The
situation is similar for the case when either, or both, of the branes
contain gauge fields, providing the D$(p-2)$-brane that remains after
the decay is aligned parallel to the initial gauge
fields~\cite{Nelson:2008eu}.

If a D$q$-brane dissolves into a D$p$-brane, which subsequently decays
due to a collision with an anti-${\bar{\rm D}}p$-brane, the gauge
fields on the D$q$-brane should allow --- at least two --- directions
perpendicular, to which the vortex (defect of co-dimension $2$) can
form. Thus, for this process to occur, the condition $q\leq p-2$ must
be met.

If this is not the case, then the bound state D$(p,p-1)$-brane and the
anti--${\bar{\rm D}}p$-brane will not be able to decay. However, one
of the two branes that makes up the bound D$(p,p-1)$-brane must be
non-BPS. This indicates that the bound state will also not be a
BPS-state and hence would not be energetically preferred to an unbound
D$p$- D$(p-1)$-brane system. In other words, we would not expect such a
bound state to form.

Let us relax the condition that the brane and anti-brane in the system
are parallel. Allowing the transverse scalars on one of the branes to
grow linearly along one direction, we effectively rotate that brane.
This may lead to problems at infinity, however since the tachyonic
decay is a local process, this issue can safely be
ignored\footnote{Alternatively one can think on transverse scalars
  varying linearly with one direction, until some specific, large,
  value where they drop again to zero. Thus, locally the brane would
  be rotated and at infinity it would become parallel to the
  anti-brane.}. In such a situation, the tachyon decays exactly as
before, with the restriction that the vortex defect contains the
transverse scalars of the original, rotated brane. Thus, in essence
the collision between two non-parallel branes behaves in exactly the
same manner as for parallel branes, simply resulting in a rotated
brane of co-dimension 2.

\section{Collisions of different dimension branes}
It was argued in Ref.~\cite{Durrer:2005nz} that the condition for two
D$p$-branes to generically intersect in $d$ space-time dimensions
reads
\be\label{eq:ppcoll}
 2p+1 \geq d-1~.
\ee 
The argument goes as follows: a D$p$-brane is extended in $(p+1)$
directions. Thus, two $p$-dimensional branes require $(2p+2)$
dimensions to exist, if there is no any special alignment of the
branes. For these branes to avoid a collision --- in the case that
there are no special, non-generic, alignments --- there must be at
least one additional dimension, namely $d\geq 2p+3$. Hence, one expects
to get generically collisions between D$p$-branes according to
Eq.~(\ref{eq:ppcoll}).

The above condition, Eq.~(\ref{eq:ppcoll}), is nevertheless not
sufficient to explain the space-time dimensionality. As it was shown
in Ref.~\cite{Durrer:2005nz}, only if the (equal dimensionality, $p$)
intersecting branes are unstable, {\sl evaporation} will eventually
take place leading to a population of remnant D3- and D1-branes.
Brane evaporation was shown to take place provided the bulk
coordinates are compactified on a torus and the branes align
themselves so that they intersect in a manifold of dimensionality
$(p-1)$ and reconnection can indeed take place. These conditions were
shown in Ref.~\cite{Durrer:2005nz} to be satisfied.  Certainly, if
bound states are not formed, then all arguments of
Ref.~\cite{Durrer:2005nz} hold, so we do not repeat them here. If the
colliding branes have unequal dimensionality, then either
statistically speaking D3-branes survive unaffected from other
dimensionality branes, since they will only intersect with D7- and
D9-branes which are the ones to evaporate first~\cite{Durrer:2005nz},
or they will form bound states.  It is the latter case that is addressed below.

The condition Eq.~(\ref{eq:ppcoll}) readily extends to collisions
between D$p$- and D$q$-branes, where for a generic intersection we
require \be p+q+1 \geq d-1~.  \ee For $d=10$ and $p\geq q $ one can
easily check that collisions will only occur for $p\geq (d-2)/2$.  In
Table~\ref{tab1} we explicitly give all possible collisions between
D$p$- and D$q$-branes, with $p\geq q$. Clearly the cases with $q \geq
p$ are obtained by simply interchanging $p$ and $q$.
\begin{table}
\begin{tabular}{|c|l|}
\hline
$p=9$\ \ & \ \ \ $q=1,3,5,7,9$ \\
$p=7$\ \ & \ \ \ $q=1,3,5,7$ \\
$p=5$\ \ & \ \ \ $q=3,5$ \\
\hline
\end{tabular}
\caption{\label{tab1} A D$p$-brane will generically collide with a D$q$-brane,
with $p\geq q$ according to this table.}
\end{table}

As we have seen in the previous sections, the collision between, for
example, a D$3$-brane and a D$5$-brane will result in a bound
D$(5,3)$-brane system, that is essentially a D$5$-brane with
(three-dimensional) world-volume gauge fields.  The collision of such
a bound state with an anti-${\bar{\rm D}}5$-brane releases the
original D$3$-brane, in the process of brane annihilation.

If we introduce the notation that a ${\rm D}^*p$-brane is either a
D$p$-brane or a bound D$(p,q)$-brane, then the collision between a
${\rm D}^*p$-brane and a ${\rm D}^*q$-brane results to a (different)
${\rm D}^*p$-brane, whilst the collision between a ${\rm D}^*p$-brane
and an anti-${\bar{\rm D}}p$-brane results is a ${\rm
  D}^*(p-2)$-brane. Also we have seen that a non-BPS ${\rm
  D}^*p$-brane will decay into a ${\rm D}^*(p-1)$-brane. In this
notation, it is only collisions between similar dimensional branes,
or their self decay due to being non-BPS, that results in lower
dimension branes.  One important feature of Table~\ref{tab1} is
that a D$3$-brane does not generically collide with a lower
dimension brane and hence ${\rm D}^*3$-branes are equivalent to
D$3$-branes.

Thus, at the end of a possibly complicated decay chain, the only
possible scenario is that a ${\rm D}^*5$-brane colliding with an
anti-${\bar {\rm D}}^*5$-brane forms a D$3$-brane, which possibly had
previously been absorbed by one of the D$5$-branes.  In conclusion,
within the context of type IIB string theory, only D$3$, and possibly
D$1$-branes, would be created from the decay of higher dimensional
branes.

We note here that although we have pointed out that D$3$-branes do
not, generically collide with lower dimensional branes, it is still
possible that bound state D$(3,q)$-branes, with $q<3$ can form. If a
bound D$(5,q)$-brane were to collide with an anti ${\bar{\rm
    D}}5$-brane, the result would be a D$(3,q)$-brane.  Such a process
allows for embedded lower dimensional branes to be present in our
universe as relics of earlier, higher dimensional collisions.  Their
existence does not alter our result, it just allows for
three-dimensional branes having absorbed one-dimensional ones.

\section{Conclusion}
We have shown that the collision and subsequent decay of higher
dimensional branes generically leads to D$3$-branes (and
anti-${\bar{\rm D}}3$-branes).  The production of (anti-) D$1$-branes
cannot be completely excluded, however such branes are formed only
through the annihilation of a pair of non-BPS branes, which would
likely be suppressed. The formation of bound states during the
collision process does not alter this result, but it just opens up the
possibility that embedded branes may be present on the
three-dimensional ones, one of which can play the r\^ole of our
universe.

\vskip.5truecm 
The work of M.S. is partially supported by the European Union through
the Marie Curie Research and Training Network {\sl UniverseNet}
(MRTN-CT-2006-035863).

\end{document}